\documentclass{PoS}
\newcommand{\cd}{\makebox[0.08cm]{$\cdot$}}
\newcommand{\sla}{\not\!}
\title{Nonperturbative calculation of the anomalous magnetic moment
in the Yukawa model}
\ShortTitle{Nonperturbative calculation of the anomalous magnetic
moment in the Yukawa model}
\author{\speaker{J.-F. Mathiot}\\
Laboratoire de Physique Corpusculaire, Universit\'e Blaise Pascal,
63177 Aubi\`ere Cedex, France\\
E-mail: \email{mathiot@in2p3.fr}}
\author{V.A. Karmanov\\
Lebedev Physical Institute, Leninsky Prospekt 53, 119991 Moscow,
Russia}
\author{A.V. Smirnov\\
Lebedev Physical Institute, Leninsky Prospekt 53, 119991 Moscow,
Russia}
\abstract{Within the covariant formulation of light-front
dynamics, we calculate the state vector of a fermion coupled to
identical scalar bosons (the Yukawa model). The state vector is decomposed in
Fock sectors and we consider the first three ones: a single
fermion, a fermion coupled to one boson, and a fermion coupled to
two bosons. This last three-body sector generates nontrivial and nonperturbative
contributions to the state vector, and these contributions are
calculated with no approximations. The divergences of the
amplitudes are regularized using Pauli-Villars fermion and boson
fields. Physical observables can be unambiguously deduced using a
systematic renormalization scheme we developed. This
renormalization scheme is a necessary condition in order to avoid uncancelled
divergences when Fock space is truncated. As an example, we
present preliminary
 numerical results for the anomalous magnetic moment of a
fermion in the Yukawa model.}
\FullConference{LIGHT CONE 2008 Relativistic Nuclear and Particle Physics\\
         July 7-11  2008\\
         Mulhouse, France}
\begin{document}
\section{Bound state systems in light-front dynamics}
\subsection{Fock representation of the state vector}
\begin{figure}[btph]
\begin{center}
\includegraphics[width=12pc]{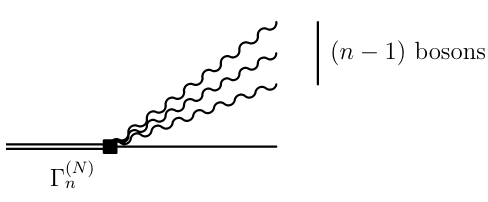}
\caption{Vertex function of order $n$ for the $N$-body Fock space
truncation.\label{gamman}}
\end{center}
\end{figure}

Light-front dynamics  enables a very convenient
representation of the state vector, $\phi (p)$, for any
relativistic bound state system with total four-momentum $p$.
In the standard formulation of light-front dynamics, the state vector is defined on
a given light front plane $t+z/c=0$. It is solution of the
eigenvalue equation $\hat{P}^{2}\ \phi(p)=M^2\ \phi(p)$, where
$\hat P$ is the momentum operator and $M$ is the bound state mass.
One of the main advantages of light-front dynamics is that, due to kinematical
constraints, the vacuum state of a system of interacting particles
coincides with the free vacuum, and all intermediate states result
from fluctuations of the physical system. One can thus construct
the state vector in terms of combinations of free fields, i.e.
decompose it in a series of Fock sectors: $ \phi(p) \equiv \vert 1
\rangle + \vert 2 \rangle +
 \dots + \vert N \rangle + \dots$
with its subsequent truncation. 
This enables a systematic calculation of state vectors of physical
systems and their observables.
We call
$N$ the maximal number of Fock sectors considered in a given
approximation, and $n$ the number of constituents in a given Fock
sector described  by the many-body vertex function
$\Gamma_n$, shown in Fig.~\ref{gamman}.

\subsection{Strict control on explicit violation of rotational invariance}
The standard formulation of light-front dynamics has however a serious drawback,
since the equation of the light front plane is not invariant under
spatial rotations. To avoid such an unpleasant feature, we use the
Covariant formulation of Light-Front Dynamics (CLFD)~\cite{cdkm}, which provides a
very powerful tool in order to describe
physical systems. In
this formulation, the state vector is defined on the plane
determined by  the equation $\omega\cd x = 0$, where
$\omega$ is an arbitrary light-like four-vector. The covariance of
our approach is due to the invariance of the light front plane. 
This implies that $\omega$ is not the
same in any reference frame, but varies according to Lorentz transformation, like the coordinate $x$. It is not the case in the standard
formulation where $\omega$ is fixed to $\omega=(1,0,0,-1)$.

This scheme is very convenient in order  to
parameterize the general spin structure of
vertex functions. For a spin-1/2 system consisting of one
fermion (with momentum $k_1$) and scalar bosons  (with momenta $k_2$, $k_3$,...),
the two- and three-body vertex functions are represented as:
\begin{equation}
\label{state}
\bar{u}(k_{1})\Gamma_{2}^{(N)}u(p)  =
\bar{u}(k_{1})\left[b_{1}+b_{2} \frac{m{ \sla \omega}}{\omega\cd p}\right]u(p)\ ,
\end{equation}
\begin{equation}
\label{state3} \bar{u}(k_{1})\Gamma_{3}^{(N)}u(p)  =
\bar{u}(k_{1})\left[c_{1}+c_{2}\frac{m{ \sla \omega}}{\omega\cd
p}+C_{ps} \left( c_{3}+c_{4}\frac{m{ \sla \omega}}{\omega\cd
p}\right) \gamma_5 \right]u(p)\ ,
\end{equation}
with
\begin{equation}
C_{ps}=\frac{1}{m^2 \omega \cd p} \ e^{\mu \nu \rho \lambda}\ k_{2\mu}\ k_{3\nu}\ p_{\rho}\ \omega_{\lambda}\ .
\end{equation}
We identified the bound state mass $M$ with the physical fermion
mass $m$. Here $b_{1,2}$ and $c_{1-4}$ are scalar functions
determined by dynamics. Generally speaking, their explicit form
depends on $N$, but the total number of irreducible spin
components for each vertex function does not. The two-body vertex
function has two, while the three-body one (as well as any of
higher order) has four irreducible components. The one-body vertex
function is just a constant which also depends on $N$.

In order to impart
sense to divergent amplitudes, we choose here the
Pauli-Villars (PV) regularization scheme which preserves
rotational invariance \cite{kms_07} and extends the state
vector to include both fermion and boson PV particles with very
large masses. An alternative scheme, based on the use of specific
test functions on which operator fields are defined, has been
developed very recently \cite{gw_07}. Its application to CLFD is
currently under study \cite{gm_08}.
\subsection{Fock sector dependent renormalization}
\begin{figure}[btph]
\begin{center}
\includegraphics[width=25pc]{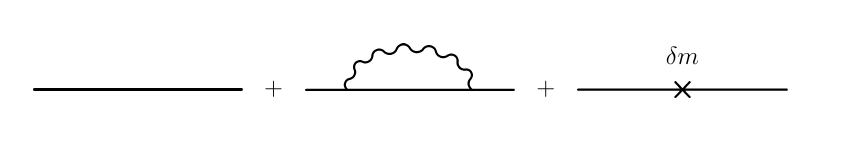}
\caption{Renormalization of the fermion propagator in second
order perturbation theory.\label{self}}
\end{center}
\end{figure}

In order to be able to make definite predictions for physical
observables, one should also define a proper renormalization
scheme. This should be done with care since the Fock decomposition
of the state vector is  truncated to a given order. Indeed,
looking at Fig. \ref{self} for the calculation of the fermion
propagator in second order
perturbation theory, one immediately realizes that the
cancellation of divergences (or terms infinitely increasing as the
PV masses tend to infinity) between the self-energy contribution
(of order two in the Fock decomposition) and the fermion Mass
Counterterm (MC) (of order one) involves two different Fock
sectors \cite{kms_08}.
This means that any MC and, more generally, any Bare Coupling
Constant (BCC) should be associated  with the number of particles
present (or ``in flight'') in a given Fock sector. In other words,
all MC's and BCC's must depend on the Fock sector under
consideration. The original MC $\delta m$ and the
fermion-boson BCC $g_0$ should thus be extended to a whole
series:
\begin{equation}
\label{g0i} g_0 \to  g_0^{(i)}\ ,
\end{equation}
\begin{equation}
\label{dmi} \delta m  \to \delta m^{(i)} \ ,
\end{equation}
with $i=1,2,\ldots N$. The quantities $g_0^{(i)}$ and
$\delta m^{(i)}$ are calculated by
solving the systems of equations for the vertex functions in
the $N=1$, $N=2$, $N=3$, ... approximations successively.
We shall illustrate this
procedure in Section 2. The
BCC $g_0^{(N)}$  is determined by demanding that the
$\omega$-independent part of the two-body vertex function $\Gamma_2$ at $s\equiv
(k_1+k_2)^2=m^2$
coincides with the physical coupling constant $g$:
\begin{equation}
\label{bare} b_{1}(s=m^2)  \equiv g\ .
\end{equation}
Note that in perturbation theory, we also have to consider a whole
series of bare parameters/counter-terms $g_0^{(n)}$ and $\delta
m^{(n)}$, where 
$n$ denotes the order of the perturbative expansion. In
light-front dynamics, the index $n$ 
refers  to the number of particles in ``flight''. A
calculation of order $N$ involves $\delta m^{(1)}\ldots \delta
m^{(N)}$ and $g_0^{(1)} \ldots g_0^{(N)}$. This procedure, which we shall call Fock Sector Dependent
Renormalization (FSDR), is a well defined, systematic, and
nonperturbative scheme \cite{kms_08}. The main difference of
our procedure from that used in \cite{bhm_06} 
consists in the use of CLFD and FSDR scheme.
\section{The anomalous magnetic moment}
\subsection{QED in two-body truncated Fock space}
The decomposition of the spin-1/2 electromagnetic
vertex  in CLFD  is given by \cite{km}:
\begin{equation}
\label{F12} \bar{u}(p')G^{\rho}u(p)=
e\bar{u}(p')\left[F_1\gamma^{\rho}+
\frac{iF_2}{2m}\sigma^{\rho\nu}q_{\nu} +B_1\left( \frac{{\sla
\omega}}{\omega\cd
p}P^{\rho}-2\gamma^{\rho}\right)+B_2\frac{m{\omega^{\rho}}}{\omega\cd
p}+B_3\frac{m^2{\sla \omega}\omega^{\rho}}{(\omega\cd
p)^2}\right]u(p)\ ,
\end{equation}
with $P=p'+p'$,  $q=p'-p$.  $F_1$ and $F_2$ are the physical form
factors, while  $B_{1,2,3}$ are spurious (nonphysical)
contributions which appear if rotational invariance is broken,
e.~g. by Fock space truncation. The decomposition (\ref{F12})
enables to separate unambiguously the physical form factors from
the nonphysical ones. Under the condition $\omega\cd q=0$, all
$F_{1,2}$, $B_{1-3}$ depend on $Q^2\equiv -q^2$ only.

The simplest realistic physical system one can
consider first is QED in two-body truncated Fock space. This
approximation is equivalent to the summation, to all orders, of
the second order perturbative correction to the electron
self-energy.
\begin{figure}[btph]
\begin{center}
\includegraphics[width=26 pc]{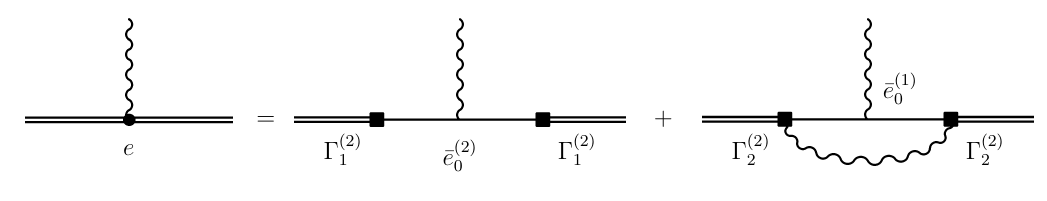}
\caption{Electromagnetic vertex of the electron in the two-body
approximation. \label{ff2} }
\end{center}
\end{figure}

The electromagnetic form factors of the electron are given by
the one- and two-body contributions shown in Fig. \ref{ff2}, where
we have denoted by $\bar e_0^{(1)}$ and $\bar e_0^{(2)}$ the
electron BCC's  which, according to our FSDR scheme, depend on the
Fock sector. Note that these BCC's describing the interaction of
an electron with an external photon field do also differ from the
"internal" photon-electron BCC's, denoted by $e_0^{(i)}$, which
appear in the calculation of the state vector itself
\cite{kms_08}, since the external photon does not participate to
the internal structure of the state vector. The calculation of the
anomalous magnetic moment  of the electron,
given by $F_2(Q^2=0)$, is thus done according to the following steps:
\begin{figure}[btph]
\begin{center}
\includegraphics[width=28 pc]{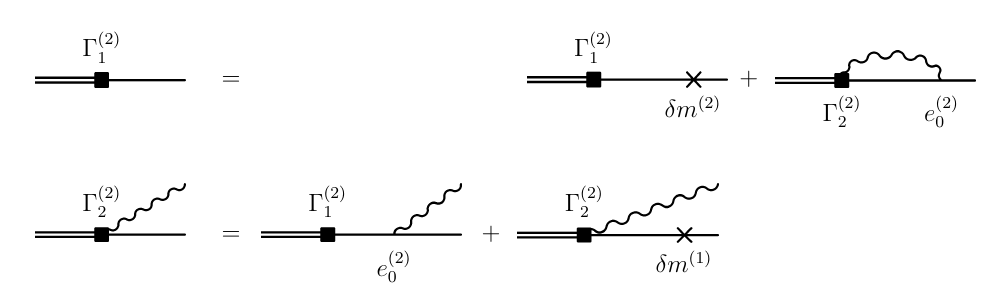}
\caption{System of equations for the vertex functions in QED for
the two-body Fock space truncation.  \label{twobody}}
\end{center}
\end{figure}
\begin{itemize}
\item We first decompose the two-body vertex function in
independent spin structures in a way very similar to
(\ref{state}), (\ref{state3}), including vector indices for the
photon line. For the $N=2$ truncation, the components $b_1$ and
$b_2$ are constants. 
\item We solve the eigenvalue equation which
is represented graphically in Fig. \ref{twobody}. Note the
appearance in this figure of the Fock sector dependent MC's and
BCC's. 
\item We determine the MC's and BCC's according to our FSDR
scheme. We have from the very beginning $\delta m^{(1)}=0$, 
while  $\delta m^{(2)}$ is fixed from the compatibility
condition of this system of two homogeneous equations 
and  $e_0^{(2)}$ is fixed from the condition analogous
to (\ref{bare}). 
\item Once the state vector is known, we
calculate the electromagnetic form factors and demand that
$F_1(Q^2=0)=1$. \ This defines $\bar e_0^{(2)}$, while $\bar
e_0^{(1)}$ is equal to $e$, since it corresponds, by definition,
to an external photon coupling to a single electron,
with no radiative corrections at all. Because of the normalization
of the state vector, and the 
counterterms which depend explicitly on the Fock sector, 
we find $\bar e_0^{(2)}\equiv e$, as dictated by the Ward
identity.
\end{itemize}

We can thus predict analytically the value of $F_2$ without any
perturbative expansion and find, in the limit of infinite PV
particle masses, $F_2(Q^2=0)=\frac{\alpha}{2\pi}$ which
coincides with the well-known perturbative Schwinger correction.
\subsection{The Yukawa model in three-body truncated Fock space}
In order to address the calculation of a true nonperturbative
system, we investigate the system composed of a fermion coupled to
scalar bosons for the three-body,
$N=3$, Fock space truncation. The strategy to
analyze  this system is very similar to
that outlined above for QED.
\begin{figure}[btph]
\begin{center}
\includegraphics[width=28pc]{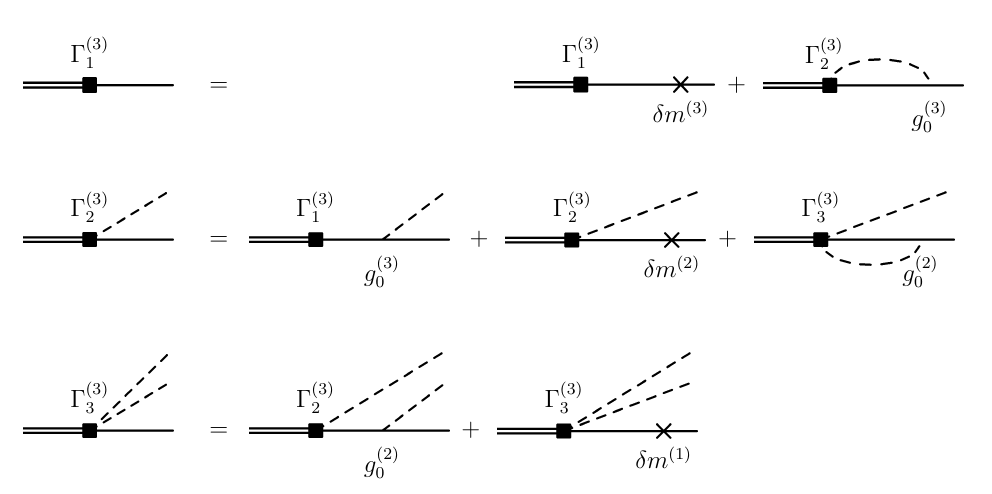}
\caption{System of equations for the
vertex functions
in the Yukawa model for the three-body Fock space truncation.
We do not show on this figure, for simplicity,  the interchange of
identical bosons in $\Gamma_3^{(3)}$. Dashed lines correspond to scalar bosons. \label{scalarfig}}
\end{center}
\end{figure}
\begin{figure}[btph]
\begin{center}
\includegraphics[width=30pc]{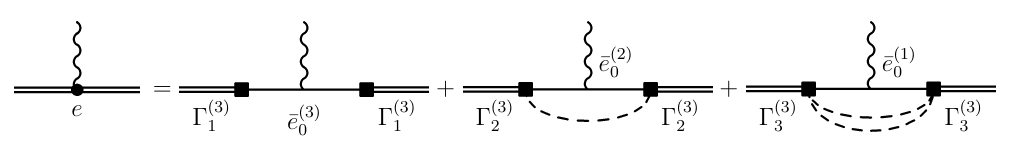}
\caption{Fermion electromagnetic vertex in the Yukawa
model for the three-body Fock space truncation. \label{WI3}}
\end{center}
\end{figure}
The system of equations one has to solve is given in Fig.
\ref{scalarfig}. Note that the indices of MC's and BCC's in this figure are
different from those present in Fig. \ref{twobody}
for the case of the two-body truncation, according to our FSDR scheme. The set of indices
just corresponds to $i=1,2,3$ in (\ref{g0i}) and (\ref{dmi}).
The electromagnetic vertex is given by Fig. \ref{WI3}.
Since the state vector is normalized, and within the FSDR scheme,
we find again $\bar e_0^{(3)}=\bar e_0^{(2)}=\bar e_0^{(1)}=e$ as
it should be. Note that the system of equations shown in Fig.
\ref{scalarfig} has a structure very similar to the one shown in
Fig. \ref{twobody} for the $N=2$ truncation. The extension to
higher order Fock space truncations is thus straightforward.

\subsection{Numerical results}

We solved numerically the system of linear integral
equations for the vertex functions, shown graphicaly in Fig.
\ref{scalarfig}, and calculated the
anomalous magnetic moment
of the fermion.
The original system of homogeneous equations 
is reduced to a system of inhomogeneous equations by
setting the one-body vertex function to a fixed value, since all $\Gamma$'s  are defined up to a normalization
constant. After discretizing the integrals by means of the
Gaussian procedure,  the solution is 
found by standard matrix
inversion methods. The vertex functions are finally normalized
\cite{kms_08}.
\begin{figure}[btph]
\begin{center}
\includegraphics[width=30pc]{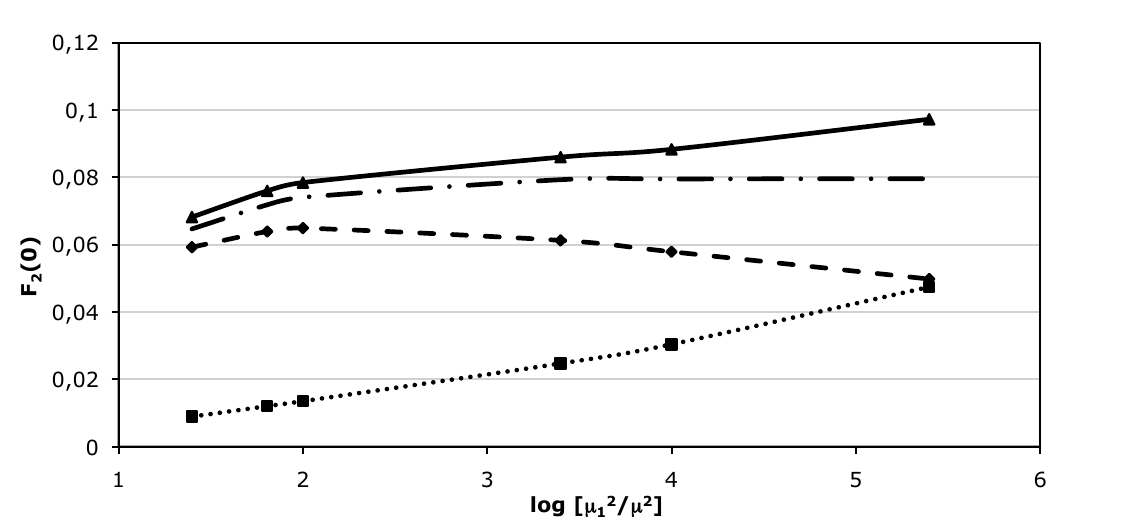}
\caption{The fermion anomalous magnetic moment as a function of
the Pauli-Villars boson mass  $\mu_1$
for the $N=3$  Fock space  truncation. We
separate the contributions from the two-body (dashed line) and
three-body (dotted line) vertex function to the total result
(solid line). The lines are just drawn to guide the eyes. The
dash-dotted line represents the anomalous magnetic moment
calculated in the N=2 
approximation.  \label{anomalous}}
\end{center}
\end{figure}

The solution depends, apart from the physical masses of the
fermion ($m$) and the boson ($\mu$) and the physical coupling
constant $g$, on the regularization parameters, namely, the masses
of the PV fermion ($m_1$) and the PV boson ($\mu_1$). After
calculating observables (say, the form factors), one should go
over to the limit of infinite PV masses. The limit
$m_1\to\infty$ can be done analytically, already at the
level of the equations for the vertex functions. This is not
 the case for the PV boson mass and we consider the value
 of the fermion anomalous magnetic moment
  as a function
of $\mu_1$,  for large values of  the latter.

For our first numerical study, we consider a typical set of
physical parameters: $ m=1$ GeV, $\mu=1$ GeV and
$\alpha=\frac{g^2}{4\pi}=1$. The results of our calculation are shown in Fig. \ref{anomalous}.
We separate on this figure the two- and three-body contributions
to the anomalous magnetic moment. The first one is slightly decreasing with
$\mu_1$ while the second is increasing. The total contribution is
rather stable, although it increases slightly with
$\mu_1$. We indicate also on this figure the fermion
anomalous magnetic moment
calculated in the  lower order approximation (i.e. in $N=2$ truncated Fock
space).
Similarly to the QED case, it has a finite limit when
$\mu_1\to\infty$. 

It remains to investigate the origin of the residual dependence of
the anomalous magnetic moment
  on $\mu_1$.
We have already shown that because of Fock state
truncation, violation of rotational invariance
may arise, leading, in particular, to a nonzero
$\omega$-dependent component in the two-body vertex function at
$s=m^2$. This nonrenormalized $b_2$, in Eq. (\ref{state}),  may contain uncancelled
$\mu_1$-dependence (even at $\mu_1\to\infty$) giving rise to
analogous dependence of the anomalous magnetic moment.
In perturbation theory, we can show  that the incorporation, in the state vector, of Fock
sectors containing fermion-antifermion pairs completely removes
extra $\omega$-dependent contributions in $\Gamma_2(s=m^2)$.
Work is in progress to extend this calculation 
to our nonperturbative approach. 

\section{Perspectives}
The general framework we have developed so far - with an
explicitly  covariant formulation of light-front dynamics and a systematic nonperturbative renormalization scheme
- enables us to calculate physical observables of
physical systems in truncated Fock space.

We have presented a preliminary  study of the fermion
anomalous magnetic moment
 in the Yukawa model,  for  a nontrivial
three-body Fock space truncation. This calculation embedded all
features (general structure of the state vector in  terms of spin
components, new nonperturbative renormalization scheme) of a more
general study and, in principle,  can be extended to the case of
Fock space truncations of arbitrary order  rather easily.
Applications to gauge theories and to effective field theories are
under consideration.

\end{document}